\documentclass[twocolumn]{book}
\usepackage[dvips]{graphicx,color}
\usepackage{makeidx,universe}

\makeauthorindex
\BookTitle{Proceedings of the XXIX PHYSICS IN COLLISION}

\CopyRight{\copyright 2009 by Universal Academy Press, Inc.}

\begin{document} %******************************************

%\tableofcontents
\pagenumbering{arabic}

\chapter{%
{\LARGE \sf
Search for Muon Neutrino Disappearance in a Short-Baseline Accelerator 
Neutrino Beam } \\
{\normalsize \bf %%%%%%%%%%%%%%******* Authors **************
Yasuhiro Nakajima, 
for the SciBooNE Collaboration } \\
{\small \it \vspace{-.5\baselineskip}%***** Affiliations ***********
Kyoto University
      Kitashirakawa-Oiwake-cho, Sakyo-ku, Kyoto 606-8502, Japan 
}
}

%**************************
% Please note:
% One \AuthorContents{} is necessary
% for EACH CONTRIBUTION, for the contents page and
% One \AuthorIndex{} is necessary
% for EACH AUTHOR, for the index.
%**************************

%***** Item below is the data for CONTENTS. 
%***** Please enter all author's name that should be initialized.
\AuthorContents{Y.\ Nakajima}

%***** items below are the data for AUTHOR INDEX. 
%***** Please enter a author's name that should be initialized.
\AuthorIndex{Nakajima}{Y.}

  \baselineskip=10pt %*******
  \parindent=10pt    %*******

\section*{Abstract} %******** Body of document starts.****************
We report a search for muon neutrino disappearance in the $\Delta m^{2}$ region of
$0.5-40 ~\rm{eV}{}^{2}$ using data from both SciBooNE and MiniBooNE experiments.
SciBooNE data provides a constraint on the neutrino flux, so that the sensitivity
to $\nu_{\mu}$ disappearance with both detectors is better than
with just MiniBooNE alone.
The preliminary sensitivity for a joint 
 $\nu_{\mu}$ disappearance search is presented.

\section{Introduction} %%%%%%%%%%%%%%%

Neutrino oscillations have been observed and confirmed at mass splitting ($\Delta m^{2}$)
of $\sim 10^{-5}~\rm{eV}^{2}$ and $\sim 10^{-3}~\rm{eV}^{2}$, called the 
``solar'' and ``atmospheric'' regions, respectively.
The observed mixing is consistent with three generations of neutrinos.

However, the LSND experiment observed an excess of $\overline{\nu}_{e}$ 
in a $\overline{\nu}_{\mu}$ beam, indicating a possible oscillation in the
$\Delta m^2 \sim 1\rm~eV^2$ region \cite{Aguilar:2001ty}.
To explain LSND with oscillations requires more than 
three generations of neutrinos or other exotic physics
beyond the Standard Model.

To test the oscillation at $\Delta m^{2} \sim 1 \rm~eV^{2}$, the MiniBooNE experiment 
recently made searches for both $\nu_{e}$ 
appearance \cite{AguilarArevalo:2007it, AguilarArevalo:2009xn}
and $\nu_{\mu}$ disappearance \cite{AguilarArevalo:2009yj}
in this parameter region.
The experiment observed no significant $\nu_{e}$ appearance signal and
ruled out as being due to 2-neutrino oscillations.
However, 
the sensitivity of MiniBooNE-only  $\nu_{\mu}$ disappearance search 
was limited by the large flux and neutrino interaction
cross-section uncertainties.

Here, we discuss an improved search for $\nu_{\mu}$ disappearance 
using data from both the SciBooNE \cite{Hiraide:2008ft} and the MiniBooNE 
experiments, where SciBooNE
detector is used to constrain flux and cross-section uncertainties.

%%%%%%%%%%%%%%%%%%%%%%%%%%%%%%%
\section{Experimental Setup}
\subsection{Fermilab Booster Neutrino Beam}

\begin{figure}[htbp]
  \begin{center}
    \includegraphics[height=5.5pc]{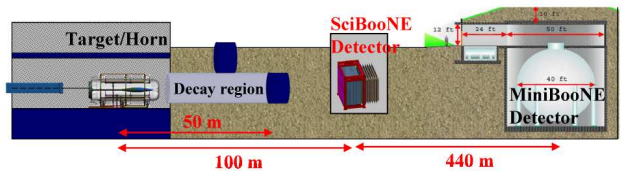}
  \end{center}
  \vspace{-1pc}
  \caption{The setup of SciBooNE and MiniBooNE experiments.}
\end{figure}

The experiments use the Booster Neutrino Beam (BNB) at Fermilab \cite{AguilarArevalo:2008yp}.
The primary proton beam,
extracted with a kinetic energy of 8 GeV, strikes a 71 cm long, 
1 cm diameter beryllium target.
The mesons, primarily $\pi^+$, generated by the $p\rm{-Be}$ interactions are focused with a magnetic horn and decay in the following 50~m decay volume, producing an intense neutrino beam with the peak energy of $\sim$0.7 GeV.
When the horn polarity is reversed,
$\pi^-$ are focused and hence a predominantly antineutrino beam is created.

\subsection{SciBooNE Detector}
The SciBooNE detector \cite{Hiraide:2008ft} is located 100 m downstream from the beryllium target.

The detector complex consists of three sub-detectors: a fully active fine grained
scintillator tracking detector (SciBar), an electromagnetic calorimeter (EC) and
a muon range detector (MRD).

The SciBar detector consists of 14,336 extruded plastic scintillator strips (CH),
each with dimension of $1.3\times 2.5\times 300$ cm$^3$.
 The scintillators are arranged vertically
and horizontally to construct a  $3\times 3\times 1.7$ m$^3$ detector. 
The detector itself is the neutrino target and its fiducial volume is 10.6 tons.

The EC is installed downstream of the SciBar,
and is made of scintillating fibers embedded in lead foil.

The MRD is located downstream of the EC in order to measure the momentum of muons up to
1.2 GeV$/c$ using the muon range. It consists of 12 layers of 2''-thick iron plates sandwiched
between layers of 6 mm-thick plastic scintillator planes.

The SciBooNE experiment ran from June 2007 until August 2008, collecting
a total of $2.52 \times 10^{20}$ Protons on Target (POT) for physics analysis;
$0.99 \times 10^{20}$ POT in neutrino mode and $1.53 \times 10^{20}$ POT in
antineutrino mode.

\subsection{MiniBooNE Detector}

The MiniBooNE detector \cite{AguilarArevalo:2008qa} 
is located 440 m downstream from the SciBooNE detector.
The detector is a 12 m diameter spherical tank filled with 
800 tons of mineral oil (CH${}_{2}$).
The MiniBooNE experiment has been taking beam data since 2002, including 
the SciBooNE and MiniBooNE joint-run period.
The collected number of POT  after data quality cut in the neutrino mode is
$5.579 \times 10^{20}$ in addition to the data from the joint-run period.

%%%%%%%%%%%%%%%%%%%%%%%%%%%%%%%

\section{$\nu_{\mu}$ Disappearance Analysis}

%%%%%%%%%%%%%%%%%%%%%%%%%%%%%%%
\subsection{Analysis Overview}
In this paper, we report only the neutrino data $(\nu_{\mu} \to \nu_{x})$ disappearance analysis.
We search for muon neutrino disappearance by comparing neutrino fluxes 
at SciBooNE and MiniBooNE detectors.

%%by comaring

The analysis is performed in the following three steps:
(1) Neutrino flux measurement at SciBooNE,
(2) Flux extrapolation to MiniBooNE, and
(3) Oscillation fit.

At each step, systematic errors are estimated and propagated to the
final prediction.
The majority of the flux and cross-section uncertainties cancels since the
neutrino interaction target in both detectors is effectively carbon, and 
the two detectors are on the same beam line. 

We describe these steps in detail in the following sections.

\subsection{Neutrino Flux Measurement at SciBooNE}
\label{sec:neutr-spectr-meas}

\subsubsection{Charged Current Event Selection}

For the spectrum analysis at SciBooNE, we use inclusive $\nu_{\mu}$
 charged current (CC) interactions, 
 whose signature is long muon tracks.
First, we identify muons by selecting the longest track with
 energy deposit consistent with a minimum-ionizing particle.
Second, we require the vertex of the track to be within the SciBar fiducial volume.
The events are further divided into two subsamples based on  the location of the
muon track end points: 
a ``SciBar-stopped" sample containing muons that 
  have stopped inside the SciBar detector and 
    a ``MRD-stopped" sample with muons that 
             have stopped in the MRD.
              These two samples each contain approximately 14k 
             and    20k events with mean energies of 0.8 and 1.1 GeV, respectively.

\subsubsection{Spectrum Fitting}
The neutrino spectrum at SciBooNE is extracted 
by fitting muon momentum ($P_{\mu})$
and muon angle ($\theta_{\mu}$) distributions from 
each sample.

\begin{figure}[htbp]
  \begin{center}
    \includegraphics[height=14pc]{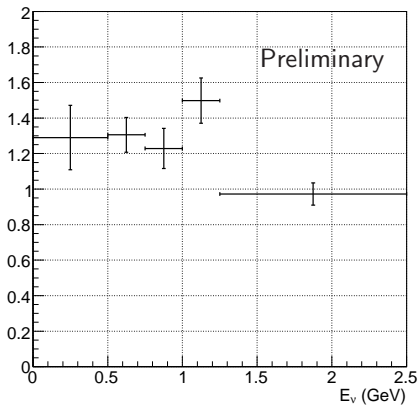}
        \begin{picture}(0,0)
   \thicklines
   \put(-80,130){ \sf Preliminary}
  \end{picture}
  \end{center}
    \vspace{-1pc}
  \caption{Scale factors obtained by SciBooNE spectrum fitting. 
  The error bars show the sum of SciBooNE statistical and systematic uncertainties.}
  \label{fig:enu_fit}
\end{figure}

We prepare MC templates for $P_{\mu}$ and $\theta_{\mu}$ distributions
for several true neutrino energy ($E_{\nu}$) regions. 
The $E_{\nu}$ regions are divided by 250 MeV up to 1.25 GeV, and a single
region is assigned for $E_{\nu} > 1.25~\rm{GeV}$.
Then,
the scale factors for each $E_{\nu}$ region are determined
to minimize the $\chi^2$ between data and MC.
Figure~\ref{fig:enu_fit} shows the fit result. The systematic errors from 
SciBooNE detector response and neutrino cross-section models are estimated 
and shown in the plot.

\begin{figure}[htbp]
  \begin{center}
        \includegraphics[height=9pc, trim = 0 9.5cm 0 0 , clip]{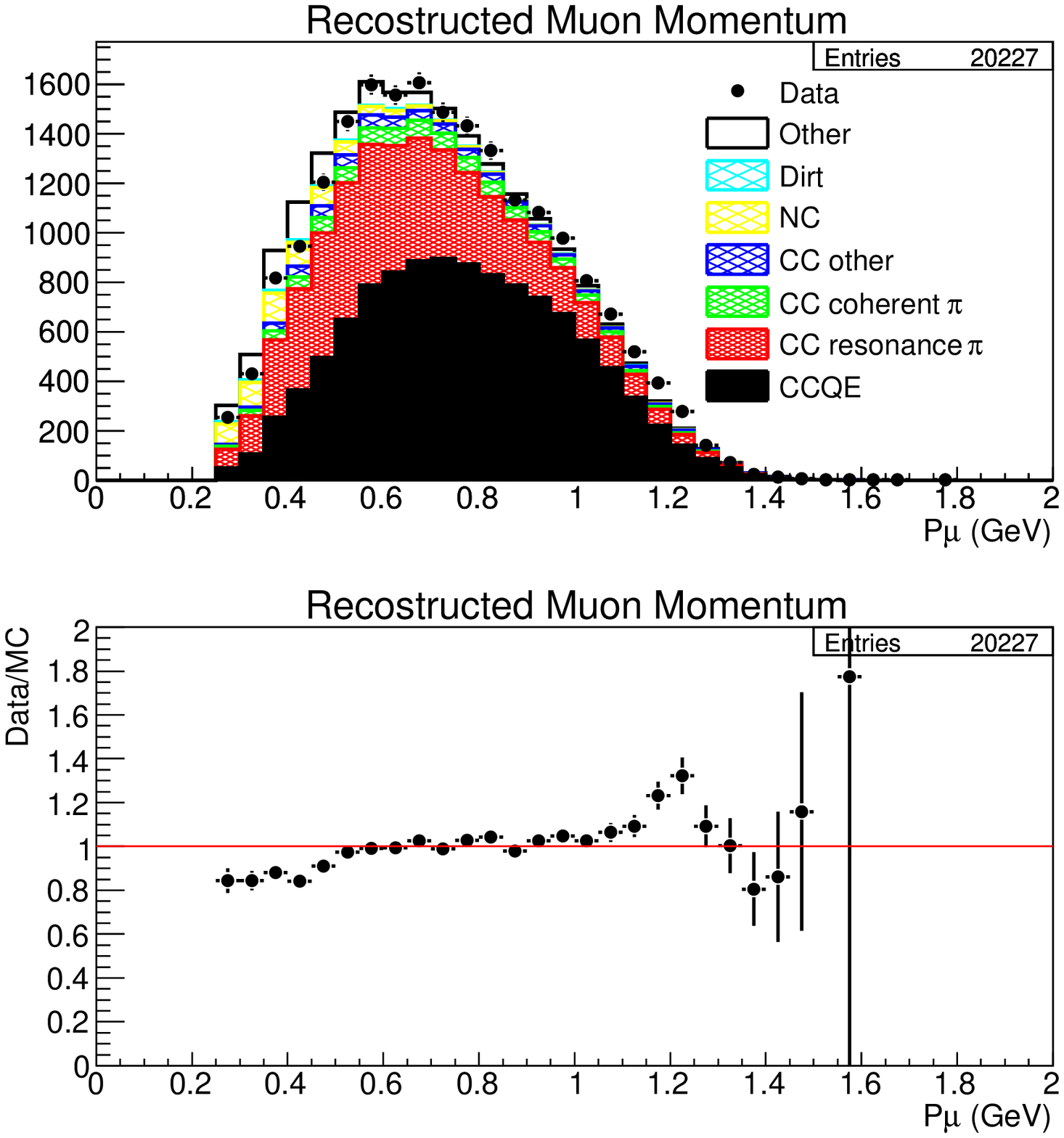}
     \includegraphics[height=9pc, trim = 0 9.5cm 0 0 , clip]{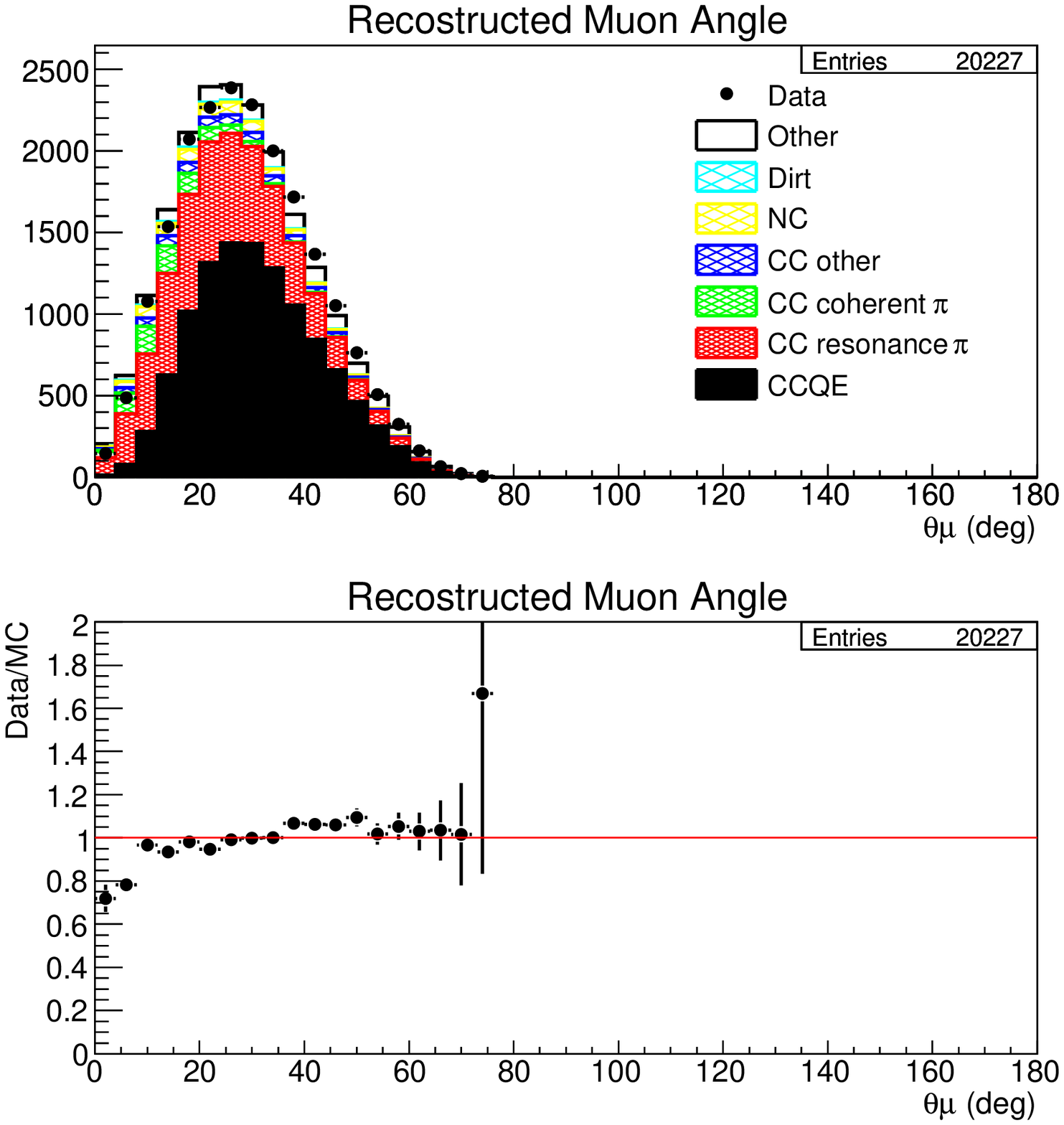}
          \begin{picture}(0,0)
	   \thicklines
	   \put(-140,195){ \sf \small Preliminary}  
   \put(-140,85){ \sf \small Preliminary}
  \end{picture}
  \end{center}
  \vspace{-1pc}
  \caption{Distribution of reconstructed muon momentum (top) and muon angle (bottom)
  for the MRD-stopped sample.
  The dots show the data, and histograms show the MC prediction with the contributions
  from neutrino interaction modes.
  The MC distributions are tuned by the E${}_{\nu}$ scale factors obtained by the
  spectrum fit.}
  \label{fig:pmu_thetamu_afterfit}
\end{figure}

Figure~\ref{fig:pmu_thetamu_afterfit} is the $P_{\mu}$ and $\theta_{\mu}$ 
distributions of SciBooNE's MRD-stopped sample, 
after applying scale factors obtained by the spectrum fitting.
We confirm the MC distributions agrees well to data after fitting.

%%%%%%%%%%%%%%%%%%%%%%%%%%%%%%%
\subsection{Flux Extrapolation to MiniBooNE}

\subsubsection{MiniBooNE Event Selection}

We select events in  MiniBooNE by 
requiring single muon and its decay electron.
Neutrino energy is reconstructed from muon kinematics 
by assuming CC Quasi Elastic (CCQE) interaction
($\nu_{\mu} n \to \mu^{-} p$):
\[
E_{\nu}^{Rec} = \frac{2(M_{n} - E_{B})E_{\mu} - (E_{B}^{2} - 2 M_{n} E_{B} 
+ \Delta M + M_{\mu}^{2})}
{2[(M_{n} -E_{B}) - E_{\mu} + p_{\mu} \cos \theta_{\mu}]},
\]
where $\Delta M = M_{n}^{2} - M_{p}^{2}$; $M$ indicate the muon, proton, or neutron mass
with appropriate subscripts; $E_{B}$ is the nucleon binding energy; $E_{\mu}$ is the 
reconstructed muon energy.

\subsubsection{MiniBooNE $E_{\nu}^{Rec}$ prediction}
To predict the $E_{\nu}^{Rec}$ distribution at MiniBooNE, we extrapolate the 
measured SciBooNE flux to MiniBooNE in two steps.

First, we apply 
MiniBooNE/SciBooNE flux ratio to make a prediction of 
the true neutrino energy distribution at MiniBooNE.
Then, we smear the true neutrino energy prediction to 
the reconstructed neutrino energy.

Systematic uncertainties for the flux ratio is estimated 
by varying the cross-section and flux models.
Additionally, the uncertainties of the smearing function, which convert
true $E_{\nu}$ to $E_{\nu}^{Rec}$, is estimated by varying the cross-section models.

Finally, we add MiniBooNE detector response error to the $E_{\nu}^{Rec}$ prediction.

\begin{figure}[htbp]
  \begin{center}
    \includegraphics[height=14pc]{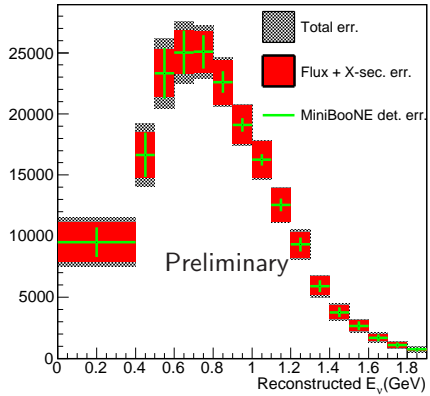}
          \begin{picture}(0,0)
   \thicklines
   \put(-120,50){ \sf Preliminary}
  \end{picture}
  \end{center}
  \vspace{-1pc}
  \caption{Predicted MiniBooNE reconstructed neutrino energy 
  distribution.
MiniBooNE detector  error, flux and cross-section uncertainty, 
and the total systematic uncertainty are separately shown.}
\label{fig:miniboone_enuqe_prediction}
\end{figure}

The predicted MiniBooNE reconstructed neutrino energy distribution and its 
systematic uncertainties are shown in the Figure~\ref{fig:miniboone_enuqe_prediction}.

%%%%%%%%%%%%%%%%%%%%%%%%%%%%%%%
\subsection{Oscillation Fit and Sensitivity}
\subsubsection{Fit Method}
We test the oscillation hypothesis assuming the mixing 
between 2 neutrino flavors; $\nu_{\mu}$ and $\nu_{x}$.
The $\nu_{\mu} \to \nu_{x}$ disappearance probability 
 is given as
 \[
P(\nu_{\mu} \to \nu_{x}) 
=  \sin^{2}2\theta \sin^{2} ( 1.27 \Delta m^{2} L /E ),
\]
where $\theta$ is the mixing angle, $\Delta m^{2}[\rm{eV}^{2}]$ is
the mass splitting between 2 flavors,  L[km] is the distance traveled and 
E[GeV] is the neutrino energy.

We fit the MiniBooNE $E_{\nu}^{Rec}$ distribution to find the best fit parameter 
minimizing the $\chi^{2}$ value:
\[
\chi^{2} = \sum^{16~bins}_{i,j} (N^{data}_{i} - N^{p}_{i} ) M^{-1}_{ij}  (N^{data}_{j} - N^{p}_{j} ),
\]
where $i,j$ denote $E_{\nu}^{Rec}$ bins, $N_{i,j}^{data}$ and $N_{i,j}^{p}$ denote
observed and predicted number of events at each bin, respectively, and
$M_{ij}$ represents statistical and systematic uncertainties 
for the final $E_{\nu}^{Rec}$ prediction.

Then we define the allowed region 
by $\Delta \chi^{2} = \chi^{2}(true) - \chi^{2}(best)$ values, where $\chi^{2}(true)$ is the
$\chi^{2}$ at the oscillation prediction being tested,
 and $\chi^{2} (best)$ is the 
smallest $\chi^{2}$ value across the $(\Delta m^{2}, \sin^{2}2 \theta)$ plane.

To obtain the confidence level at each oscillation parameter 
point ($\Delta m^{2}, \sin^{2}2 \theta$), we use Feldman-Cousins' 
method \cite{Feldman:1997qc}.
In this method, 1000 ``fake-data'' predictions are formed, 
using random draws of the  statistical 
and systematic uncertainties and some underlying oscillation hypothesis.
Then, each fake-data is fit  to obtain the relation between 
the $\Delta \chi^{2}$ values and the corresponding probabilities.
This process is repeated for each pair of  ($\Delta m^{2}, \sin^{2}2 \theta$) 
true oscillation parameter being tested.

\subsubsection{Expected Limit}
The sensitivity is defined as the average of limits obtained from fake experiments
with null underlying oscillation.

\begin{figure}[htbp]
  \begin{center}
    \includegraphics[height=14pc]{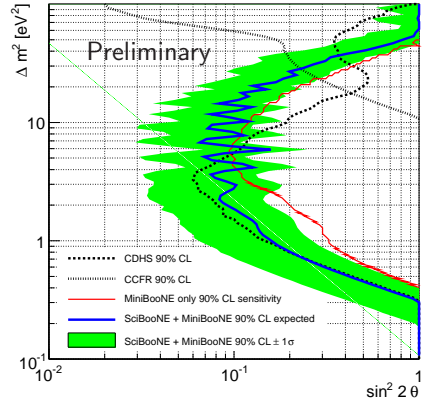}
      \begin{picture}(0,0)
   \thicklines
   \put(-150,130){ \sf Preliminary}
  \end{picture}
  \end{center}
  \vspace{-1pc}
  \caption{The expected sensitivity for $\nu_{\mu}$ disappearance.
  The dotted curve shows the 90\% CL limits from CDHS~\cite{Dydak:1983zq}
   and CCFR~\cite{Stockdale:1984cg} experiments.
  The thin solid  curve is the MiniBooNE-only 90\% CL sensitivity. 
  The thick solid curve and the filled region are the 90\% CL sensitivity 
  and $\pm 1 \sigma $ band   from SciBooNE-MiniBooNE joint analysis, respectively. }
  \label{fig:sensitivity}
\end{figure}

Figure~\ref{fig:sensitivity} shows the 90\% CL. sensitivity for 
the $\nu_{\mu}$ disappearance. The expected $\pm 1 \sigma$  band is also shown 
in the plot.
The expected sensitivity directly supersedes the MiniBooNE only
$\nu_{\mu}$ disappearance result, as substantial flux and cross section
uncertainties have been reduced.

\section{Summary and Prospects}
We present SciBooNE-MiniBooNE joint analysis of a search for 
$\nu_{\mu}$ disappearance in a accelerator neutrino beam.
The analysis is sensitive to the oscillation at the $\Delta m^{2}$ region of
$0.5 - 40 ~\rm{eV}^{2}$.
The sensitivity to $\nu_{\mu}$ disappearance has been improved relative to
the MiniBooNE shape-only analysis, with results to be released soon.
In addition, a joint  anti-neutrino oscillation analysis will be performed 
using the anti-neutrino data set.

\section{Acknowledgements}
SciBooNE collaboration gratefully acknowledges the support from various
grants and contracts from the Department of Energy (U.S.), the National
Science Foundation (U.S.), the MEXT (Japan), the INFN (Italy) 
the Ministry of Education and Science and 
CSIC (Spain), and the STFC (UK). 
We thank MiniBooNE collaboration for various informations and simulation outputs.
The author was supported by Japan Society for the Promotion of Science, and 
by the Grant-in-Aid for the Global COE Program ``The Next Generation of Physics, Spun from Universality and Emergence'' from the  MEXT of Japan.
%%%%%%%%%%%%%%%%%%%%%%%%%%%%%%%%%%
%% thebibliography environment %%
%%%%%%%%%%%%%%%%%%%%%%%%%%%%%%%%%

%%%%%%%%%%


\begin{thebibliography}{99}%%%%%%%%%%


   %%%%% LSND result
\bibitem{Aguilar:2001ty}  
	   A. Aguilar et al.,
	Phys. Rev. D64 (2001) 112007.

 %% MiniBooNE nue appearance
\bibitem{AguilarArevalo:2007it}
A. A. Aguilar-Arevalo et al.,
Phys. Rev. Lett. 98 (2007) 23180.
 
 %% MiniBooNE nue-bar appearance
\bibitem{AguilarArevalo:2009xn}
A.~A. Aguilar-Arevalo et~al.
Phys. Rev. Lett. 103 (2009) 111801.

 %% MiniBooNE  numu disappearance
\bibitem{AguilarArevalo:2009yj}
A. A. Aguilar-Arevalo et al.,
Phys. Rev. Lett. 103 (2009) 061802.


	
 %% SciBooNE CC-coh paper
\bibitem{Hiraide:2008ft}
K. Hiraide et al.,
Phys. Rev. D78 (2008) 112004.

%% MiniBooNE  Flux paper
\bibitem{AguilarArevalo:2008yp} 
A. A. Aguilar-Arevalo et al.,
Phys. Rev. D79 (2009) 072002. 


%% MiniBooNE  Detector paper
\bibitem{AguilarArevalo:2008qa} 
A. A. Aguilar-Arevalo et al.,
Nucl. Instrum. Meth. A599 (2009) 28.


 %% Feldman-Cousins
\bibitem{Feldman:1997qc} 
G. J. Feldman and R. D. Cousins.,
Phys. Rev. D57 (1998) 3873.

%% CDHS
\bibitem{Dydak:1983zq}
F. Dydak et al., Phys. Lett. B134 (1984) 281.


%% CCFR
\bibitem{Stockdale:1984cg}
I. E. Stockdale et al., Phys. Rev. Lett. 52 (1984) 1384.




 \end{thebibliography}
\end{document}